\documentclass[12pt,a4paper]{article}
\usepackage[english]{babel}
\usepackage{misccorr}
\usepackage{graphicx}
\usepackage{amsmath}

\def\eps{\varepsilon}
\def\ds{\displaystyle}

\begin{document}
\title{Modeling free anyons at the bosonic and fermionic ends}
\author{Yanina Vasiuta, Andrij Rovenchak,\\[6pt]
Department for Theoretical Physics,\\
Ivan Franko National University of Lviv,\\
12 Drahomanov St, Lviv, Ukraine}

\maketitle

\abstract{
The topology of two-dimensional movement allows for existing of anyons -- particles obeying statistics intermediate between that of bosons and fermions. In this article, the functional form of the occupation numbers of free anyons is suggested as a modification of the Gibbs factor in the Bose and Fermi statistics. The proposed expressions are studied in the bosonic and fermionic limits. The obtained virial coefficients coincide with those of free anyons up to the fourth and fifth virial coefficients (the proposed approach can be extended for higher ones as well) and up to the second order in the anyonic parameter. The effective excitation spectrum corresponding to anyons is calculated.\\[0pt]

}

\section{Introduction}
The theoretical possibility of existence of what is called `anyons' was made in 1977 \cite{Leinaas&Myrheim:1977}. In early 1980s, Wilczek proposed a model based on the (2+1)-dimensional electrodynamics and showed that an exchange of two particles leads to an arbitrary (`any') change of the wave function phase, so he proposed the term `anyons' \cite{Wilczek:1982c}.

The discovery of the fractional quantum Hall effect \cite{Tsui_etal:1982} provoked a search of quasi-particle theories, which are able to describe this phenomenon: Laughlin \cite{Laughlin:1983}, Halperin \cite{Halperin:1984}, Arovas et al. \cite{Arovas_etal:1994}. Some experiments \cite{Wang_etal:1992,Camino_etal:2005} seem to confirm the existence of fractionally charged excitations and hence indirectly of anyons. Recently, exotic fractional Hall states with $\nu=2+{}^3{\mskip -5mu/\mskip -3mu}_8$ were modeled using non-Abelian anyons \cite{Hutasoit_etal:2017}. Besides, non-Abelian Fibonacci anyons are identified with the lowest energy charged excitation in the fractional quantum Hall plateau with  $\nu={}^{12}{\mskip -5mu/\mskip -3mu}_5$ \cite{Mong_etal:2017}. Topological bubbles of Abelian anyons also can be observed in fractional quantum Hall effect \cite{Han_etal:2016}. As it can be seen, fractionally charged excitations are described in many other works, such as \cite{Braggio_etal:2016,Rech_etal:2017}, etc. 

Another place, where anyons play an important role, is high-$T_c$ superconductors \cite{Yang&Ge:1994}. A system which consists of a superconducting film on a semiconductor heterotransition gives us one more experimental observing of anyons \cite{Weeks_etal:2007}. 

Also anyons can potentially be used in quantum computing. For example, Freedman et al. \cite{Freedman_etal:2002} proved that for certain types of non-Abelian anyons \cite{Polychronakos:2000,You_etal:2013,Mancarella_etal:2013} braiding enables one to perform universal quantum computation \cite{Burton_etal:2017,Li_etal:2017}.

Error control in quantum computing is an important task \cite{Knapp_etal:2016} and such problems are closely linked to studies of various fractional statistics types. For instance, effective models of Dirac fermions and their analogies in the anyonic statistics are analyzed in \cite{Lapa_etal:2016}. Paper \cite{Gavrilik&Kachurik:2016} deals with the properties of $q$- and $p,q$-deformed oscillators.

So far, statistical mechanics cannot describe anyons completely. Because of statistical interactions between anyons even a non-interacting system is hard to be calculated. With the help of approximate approaches it is possible to find correspondences with fractional statistics types using the second virial coefficient \cite{Khare:2005}, but this method fails to catch the third one. So, the distribution function of anyons remains a puzzle.

This puzzle can be partially solved within a deformation of a usual statistics. Such an approach was made in several previous works, for example, the $q$-deformations of commutation relations between the creation and annihilation operators \cite{Dalton&Inomata:1995,Algin_etal:2002,Algin:2010,Gavrilik&Mishchenko:2013} are frequently used.

Also anyonic effects can be taken into account by deformations of a distribution function of the system. Previously it was made by Polychronakos \cite{Polychronakos:1996}, Haldane \cite{Haldane:1991} and Wu \cite{Wu:1994}. These approaches introduce one parameter into the model and hence change the form of the distribution function. Also a deformation of the exponential in the Gibbs factor can be made using the Tsallis approach \cite{Tsallis:1988} and this gives a second statistical parameter. Such two-parametric models of fractional statistics were proposed to obtain an expression for the occupation numbers of free anyons 
\cite{Rovenchak:2014EPJB,Hornetska&Rovenchak:2016en}. This approach ensures effective models for the occupation numbers of anyons exact up to the third virial coefficient. We will move further in this work considering bosonic and fermionic limits.

The paper is organized as follows. In Section~\ref{sec:modif}, a modification of statistics is reviewed and general expressions for virial coefficients of anyon gas with appropriate relations for virial and cluster expansions are considered. The expansion for occupation number in fugacity is obtained. In Sections~\ref{sec:bosonic} and \ref{sec:fermionic}, respectively, the bosonic and fermionic limits are considered. Section~\ref{sec:spectrum} is devoted to the calculations of the excitation spectrum. A short discussion in Section~\ref{sec:discussion} concludes the paper.

\section{Modification of statistics}\label{sec:modif}
The aim of this Section is to modify expressions for occupation numbers in standard quantum statistics to make them mimic the anyonic system near the bosonic and fermionic ends. So, our idea is to consider the expression for occupation numbers 
\begin{equation}\label{occ_num}
n^{\rm B,F}_j=\frac{1}{z^{-1}\exp\left(\frac{\eps_{j}}{T}\right)-\gamma}
\end{equation}
with $\gamma=\pm1$ for Bose (`B', upper sign) and Fermi (`F', lower sign), respectively, and to replace the Gibbs exponential in (\ref{occ_num}) by the following function 
\begin{equation}
\exp\left(\frac{\eps_{j}}{T}\right) \rightarrow X\left(\frac{\eps_j}{T}\right)=\exp\left(\frac{\eps_{j}}{T}\right) \left[1+\eta f_1\left(\frac{\eps_{j}}{T}\right)+\theta f_2\left(\frac{\eps_{j}}{T}\right)\right],
\end{equation}
where $\eta\sim\alpha$, $\theta\sim\alpha^2$ with $\alpha\in[0;1]$ being the anyonic parameter corresponding to the phase shift due to the permutation of particles $|21\rangle = e^{i\pi\alpha}|12\rangle$. Then the expression for the occupation numbers reads
\begin{equation}\label{occ_num2}
n_j=\frac{1}{z^{-1}X\left(\frac{\eps_{j}}{T}\right)-\gamma}.
\end{equation}
We will find the unknown functions $f_1$ and $f_2$ from the correspondence between the virial coefficients of anyons and those of a system described by the proposed fractional statistics.

Virial coefficients of the anyon gas are well-known. So, the second and third virial coefficients of an ideal anyon gas read \cite{Mashkevich_etal:1996}
\begin{align}\label{4}
&b_2^{\rm anyon}=-\frac{1}{4}(1-4\alpha+2\alpha^2),\nonumber \\
&b_3^{\rm anyon}=\frac{1}{36}+\frac{\sin^2{\alpha}}{12\pi^2}+c_3\sin^4{\pi\alpha},\\[6pt]
&\qquad\qquad c_3=-(1.652\pm0.012)\times 10^{-5}.\nonumber
\end{align}
The fourth virial coefficient equals \cite{Kristoffersen_etal:1998}:
\begin{align}\label{5}
&b_4^{\rm anyon}=\frac{\sin^2\pi\alpha}{16\pi^2}\left(\frac{\ln(\sqrt{3}+2)}{\sqrt{3}}+\cos\pi\alpha\right)
	+(c_4+d_4\cos\pi\alpha)\sin^4\pi\alpha,\\[6pt]
&\qquad\qquad
c_4=-0.0053\pm 0.0003,\quad 
d_4=-0.0048\pm 0.0009. \nonumber
\end{align}

To find virial coefficients of the gas with the distribution function given by Eq.~(\ref{occ_num2}) let us use the relation between virial coefficients and cluster integrals~\cite{Borges_etal:1999}
\begin{align}\label{6}
&b_2\lambda^2=-\frac{\mathcal{B}_2}{\mathcal{B}_1^2}, \nonumber \\[6pt]
&b_3\lambda^4=-2\frac{\mathcal{B}_3}{\mathcal{B}_1^3}+4\frac{\mathcal{B}_2^2}{\mathcal{B}_1^4},\\[6pt]
&b_4\lambda^6=-3\frac{\mathcal{B}_4}{\mathcal{B}_1^4}+18\frac{\mathcal{B}_2\mathcal{B}_3}{\mathcal{B}_1^5}-20\frac{\mathcal{B}_2^3}{\mathcal{B}_1^6}, \nonumber\\[6pt]
&b_5\lambda^8=-4\frac{\mathcal{B}_5}{\mathcal{B}_1^5}
+\frac{32 \mathcal{B}_2 \mathcal{B}_4 + 18 \mathcal{B}_3^2}{\mathcal{B}_1^6} 
- 144 \frac{\mathcal{B}_2^2 \mathcal{B}_3}{\mathcal{B}_1^7} + 112 \frac{\mathcal{B}_2^4}{\mathcal{B}_1^8}.\nonumber
\end{align}

We will consider a two-dimensional system of $N$ free particles with mass $m$ on the area ${\mathcal{A}}$ obeying distribution (\ref{occ_num2}). Cluster integrals can be found from the following expansion in fugacity $z$ \cite{Khare:2005}
\begin{equation}\label{7}
\frac{N}{\mathcal{A}}=\frac{1}{\mathcal{A}}\sum_{j}g_jn_j=\sum_{l=1}^{\infty}lB_lz^l,
\end{equation}
where the sum runs over all the energy levels with degeneracies $g_j$.
	
On the other hand, it is possible to rewrite the above expression in the integral form using the density of states $g(\eps) = \frac{Tm\mathcal{A}}{2\pi \hbar^2}={\rm const}$:
\begin{equation}\label{8}
\frac{N}{\mathcal{A}}=\frac{1}{\mathcal{A}}\sum_{j}g_jn_j=\frac{1}{\mathcal{A}}\int_{0}^{\infty}d\xi\, \frac{Tm\mathcal{A}}{2\pi \hbar^2}n(\xi)=\lambda^{-2}\int_{0}^{\infty}d\xi\, n(\xi),
\end{equation}
where $\ds\lambda=\sqrt{\frac{2\pi\hbar^2}{mT}}$ is the de~Broglie thermal wavelength, with $n(\xi)$ given by Eq.~(\ref{occ_num2}),
\begin{equation}\label{occ_num2a}
n(\xi)=\frac{1}{z^{-1}X\left(\xi\right)-\gamma}.
\end{equation}

The bosonic and fermionic limits of the modified fractional statistics are provided by $\gamma=+1$ with the anyonic parameter $\alpha\to 0$ and $\gamma=-1$ with $\alpha=1-\alpha',\ \alpha'\to 0$, respectively. Let us expand the expression for occupation numbers in fugacity:
\begin{multline}\label{9}
n(\xi)=\frac{ze^{-\xi}}{1+\eta f_1(\xi)+\theta f_2(\xi)}\Big(1\pm\frac{ze^{-\xi}}{1+\eta f_1(\xi)+\theta f_2(\xi)} {} \\ {}+ \frac{z^2e^{-2\xi}}{[1+\eta f_1(\xi)+\theta f_2(\xi)]^2}\pm\frac{z^3e^{-3\xi}}{[1+\eta f_1(\xi)+\theta f_2(\xi)]^3}+\cdots\Big),
\end{multline}
where the upper sign is for the bosonic and the lower sign is for the fermionic end.
By equating terms with identical fugacity powers in Eqs.~(\ref{7}) and (\ref{8}) after expanding expression (\ref{9}) in $\eta$ and $\theta$ one can find proper expressions for cluster integrals $\mathcal{B}_l$
\begin{align}\label{10}
&\mathcal{B}_1=\lambda^{-2}\int_{0}^{\infty}d\xi\, e^{-\xi}\left[1-\eta f_1-\theta f_2+\eta^2f_1^2\right],\nonumber \\[6pt]
&\mathcal{B}_2=\pm\lambda^{-2}\int_{0}^{\infty}d\xi\, e^{-2\xi}\left[\frac{1}{2}-\eta f_1-\theta f_2+\frac{3}{2}\eta^2f_1^2\right], \nonumber \\[6pt]
&\mathcal{B}_3=\lambda^{-2}\int_{0}^{\infty}d\xi\, e^{-3\xi}\left[\frac{1}{3}-\eta f_1-\theta f_2+2\eta^2f_1^2\right],\\[6pt]
&\mathcal{B}_4=\pm\lambda^{-2}\int_{0}^{\infty}d\xi\, e^{-4\xi}\left[\frac{1}{4}-\eta f_1-\theta f_2+\frac{5}{2}\eta^2f_1^2\right],\nonumber\\[6pt]
&\mathcal{B}_5=\lambda^{-2}\int_{0}^{\infty}d\xi\, e^{-4\xi}\left[\frac{1}{5}-\eta f_1-\theta f_2+3\eta^2f_1^2\right] \nonumber
\end{align}

For the sake of simplicity, the functions $f_1(\frac{\eps_{j}}{T})$ and $f_2(\frac{\eps_{j}}{T})$ can be chosen as follows
\begin{align}\label{eq:f_xi}
&f_j(\xi)=e^{-\xi}\sum_{\ell=1}^{\ell_{\max}}k_{j\ell}\xi^\ell,
\end{align}
where $k_{j\ell}$ are unknown constants. Such a choice significantly facilitates analytical treatment of the considered problem.

\section{Bosonic limit}\label{sec:bosonic}
Since we are interested in the anyonic behavior in the vicinity of the bosonic ($\alpha\rightarrow0$) and fermionic ($\alpha'=1-\alpha\rightarrow0$) ends, let us expand expressions (\ref{4}) and (\ref{5}) for anyonic virial coefficients in $\alpha$ or $\alpha'$, respectively. In the bosonic limit series up to $\alpha^2$ are
\begin{align}
&b_2^{\rm anyon} = -\frac14 + \alpha - \frac12 \alpha^2,\\[12pt]
&b_3^{\rm anyon} = \phantom{-}
\frac{1}{36} + \frac{1}{12} \alpha^2,\\[12pt]
&b_4^{\rm anyon} \simeq \phantom{-}0.110022 \alpha^2.
\end{align}

Expanding to the same accuracy viral coefficients from Eq.~(\ref{6}) with cluster integrals given by Eq.~({\ref{10}) one obtains a set of linear equations for $k_{j\ell}$, though rather cumbersome. While the zeroth order in $\alpha$ is satisfied automatically, we have six equations:

\begin{tabbing}
linear in $\alpha$: \quad\qquad\= $b_2^{\rm anyon} = b_2$, \qquad 
$b_3^{\rm anyon} = b_3$, \qquad
$b_4^{\rm anyon} = b_4$;\\[12pt]
quadratic in $\alpha$: \>$b_2^{\rm anyon} = b_2$, \qquad 
$b_3^{\rm anyon} = b_3$, \qquad
$b_4^{\rm anyon} = b_4$.
\end{tabbing}

\noindent
The solutions are the following values for the bosonic end:
%
\begin{align}
&k_{11} = a, \qquad
k_{12} = -0.8675\, a\qquad
k_{13} = 0.5007\, a\nonumber\\[6pt]
&k_{21} = b, \qquad
k_{22} = -0.8723\, b\qquad
k_{23} = 0.1309\, b\\[6pt]
&\textrm{with}\qquad
a = -\frac{679}{18}\simeq37.72\quad\textrm{and}\quad
b \simeq -373.0\nonumber
\end{align}
yielding correct (up to $\alpha^2$) virial coefficients $b_2$ through $b_4$.

The results of calculations are compared to the virial coefficients of anyons in Figs.~\ref{fig:b3Bose} and \ref{fig:b4Bose}.

\begin{figure}[h]
\centerline{\includegraphics[scale=0.7]{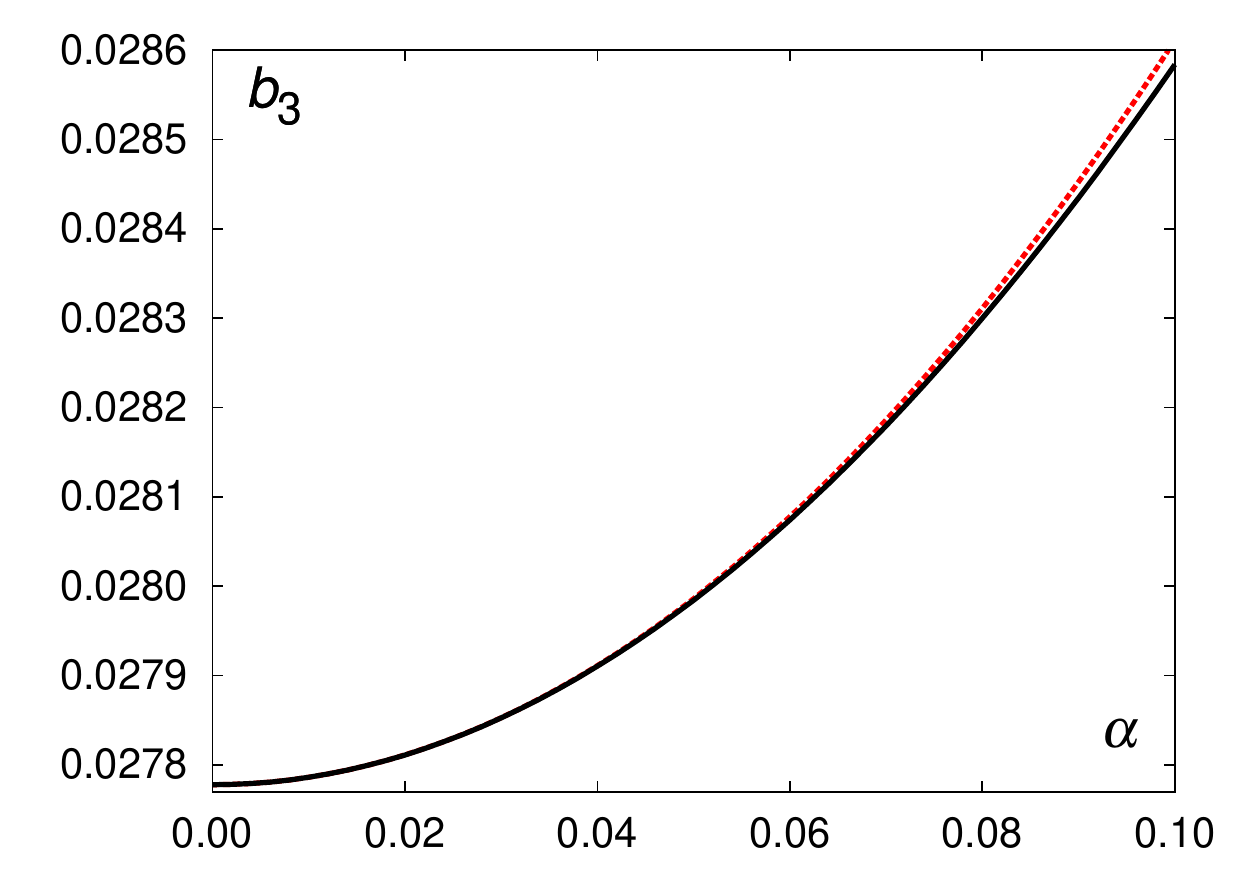}}
\centerline{\includegraphics[scale=0.7]{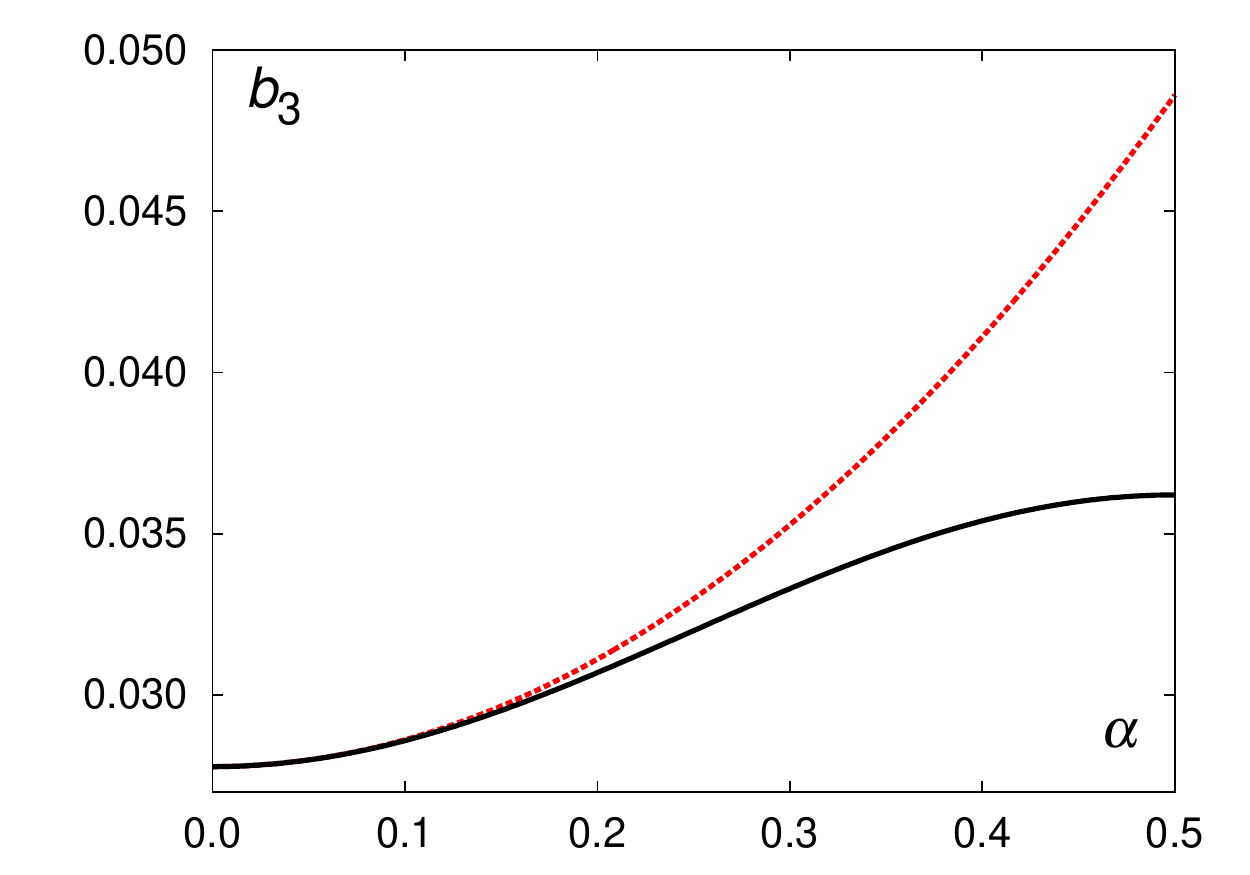}}
\caption{The third virial coefficient:
black solid line --- anyons;
red dotted line --- calculated.
The upper panel corresponds to the close vicinity of the bosonic end while the lower one demonstrates the whole bosonic side $0\leq\alpha\leq1/2$.
}\label{fig:b3Bose}
\end{figure}

\begin{figure}[h]
\centerline{\includegraphics[scale=0.7]{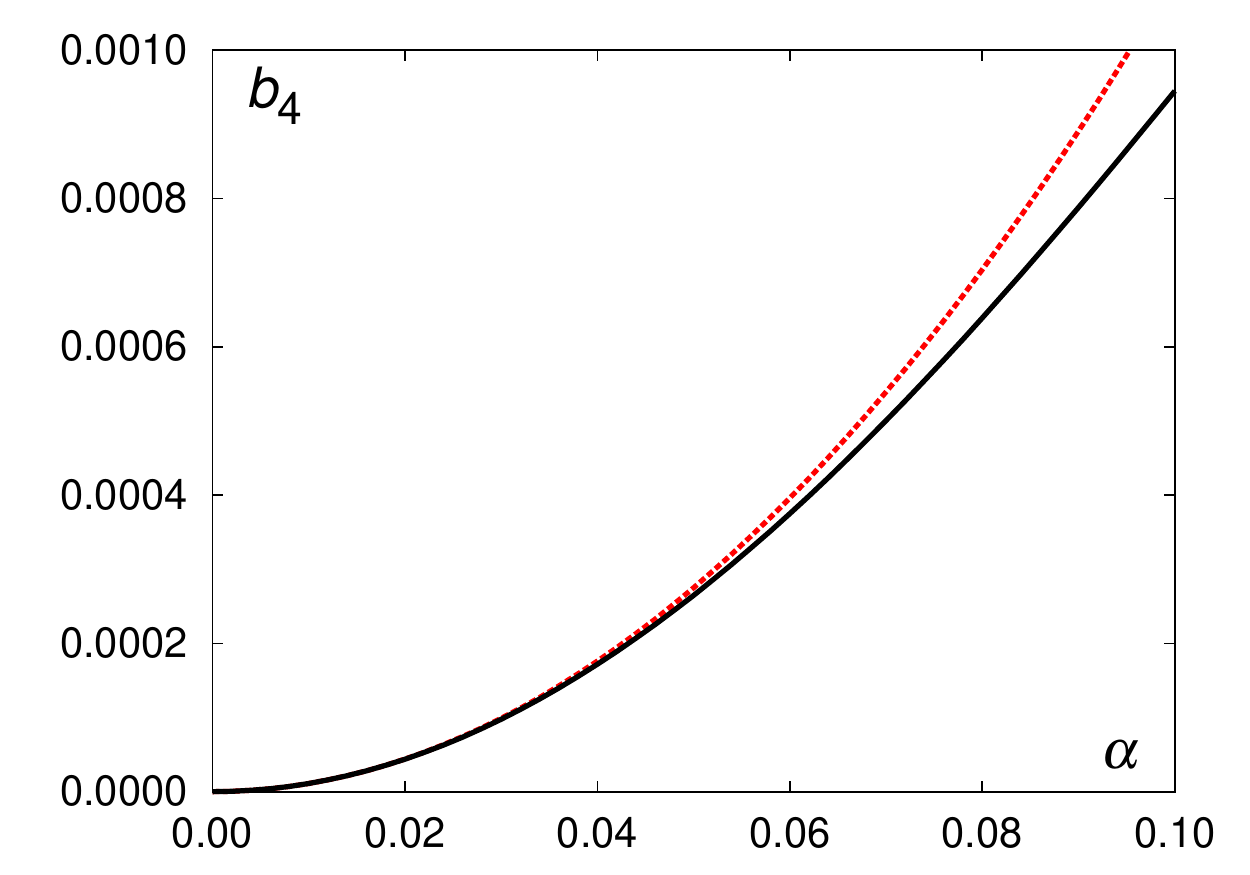}}
\centerline{\includegraphics[scale=0.7]{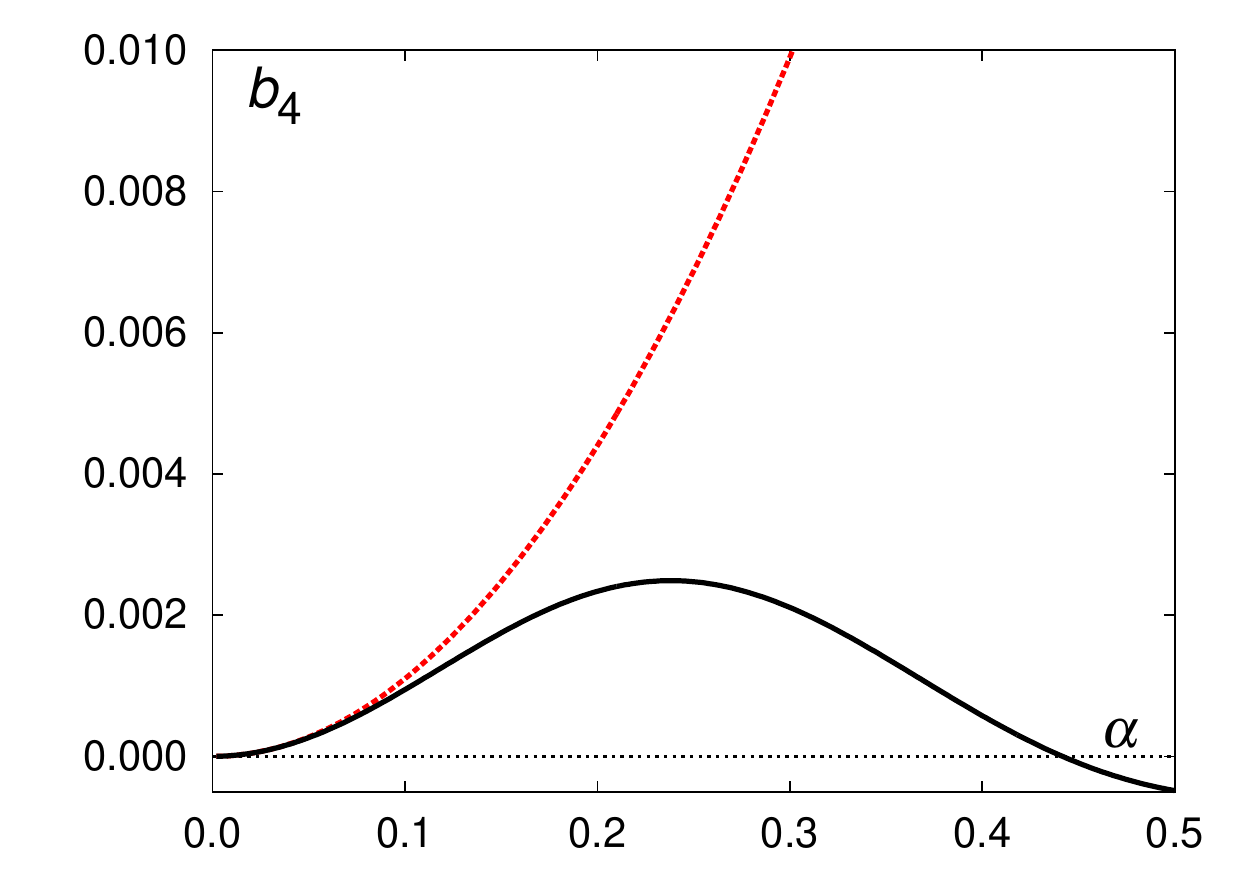}}
\caption{The fourth virial coefficient:
black solid line --- anyons;
red dotted line --- calculated.
The upper panel corresponds to the close vicinity of the bosonic end while the lower one demonstrates the whole bosonic side $0\leq\alpha\leq1/2$.
}\label{fig:b4Bose}
\end{figure}
\clearpage

A natural question arises whether the obtained results affect the values of the fifth virial coefficient. For anyons, it equals in the bosonic limit \cite{Dasnieres&Ouvry:1992}
\begin{equation}\label{eq:b5}
b_5^{\rm anyon} \simeq -\frac{1}{3600} + 0.0570337 \alpha^2.
\end{equation}
The above results reproducing the third and fourth virial coefficients yield the fifth virial coefficient shown in  Fig.~\ref{fig:b5Bose} compared to function (\ref{eq:b5}).

\begin{figure}[h]
\centerline{\includegraphics[scale=0.7]{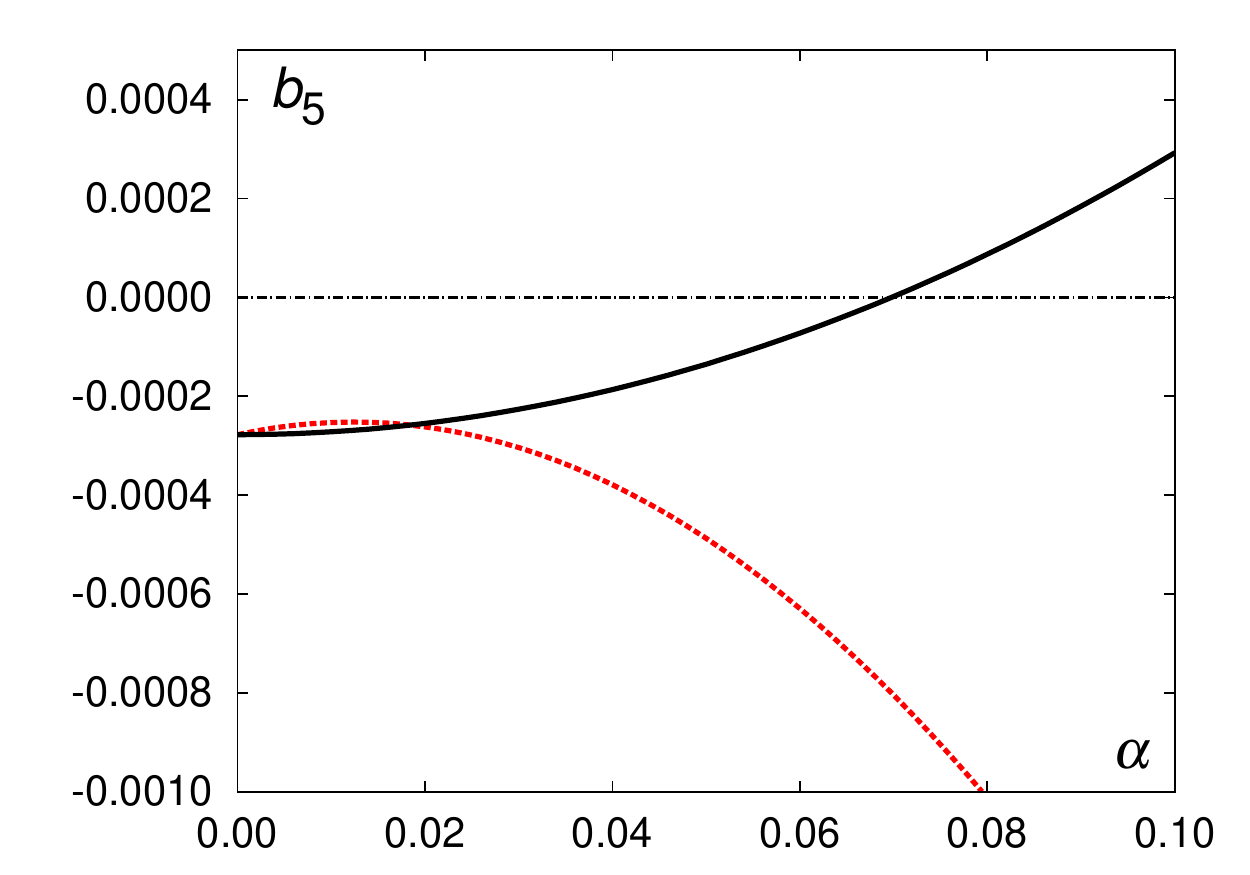}}
\caption{The fifth virial coefficient:
black solid line --- anyons;
red dotted line --- calculated.
}\label{fig:b5Bose}
\end{figure}

One can easily notice that, except for an immediate vicinity of $\alpha=0$, the anyonic and calculated $b_5$ have opposite dependences on $\alpha$, just like it was observed in \cite{Rovenchak:2014EPJB,Hornetska&Rovenchak:2016en} for the fourth virial coefficient.

It is possible to find such expressions for $f_{1,2}(\xi)$ demanding that expansions of virial coefficients $b_2$ through $b_5$ in $\alpha$ coincide with the anyonic results. In this case, $\ell_{\rm max} = 4$ and we have a set of eight equations
\begin{tabbing}
linear in $\alpha$: \qquad \= $b_2^{\rm anyon} = b_2$, \quad 
$b_3^{\rm anyon} = b_3$, \quad
$b_4^{\rm anyon} = b_4$, \quad
$b_5^{\rm anyon} = b_5$;\\[12pt]
quadratic in $\alpha$: \>$b_2^{\rm anyon} = b_2$, \quad 
$b_3^{\rm anyon} = b_3$, \quad
$b_4^{\rm anyon} = b_4$, \quad
$b_5^{\rm anyon} = b_5$.
\end{tabbing}
The solutions are as follows:
\begin{align}
&\hspace*{-0.5cm}
k_{11} = a, \quad
k_{12} = -0.6481\, a\quad
k_{13} = 0.3198\, a\quad
k_{14} = 0.04833\, a\nonumber\\[6pt]
&\hspace*{-0.5cm}
k_{21} = b, \quad
k_{22} = -1.4145\, b\quad
k_{23} = 0.7391\, b\quad
k_{24} = -0.1321\, b\\[6pt]
&\hspace*{-0.5cm}
\textrm{with}\quad
a = -\frac{71389}{2178}\simeq32.78\quad\textrm{and}\quad
b = -511.8.\nonumber
\end{align}
So, the above values yield correct (up to $\alpha^2$) virial coefficients $b_2$ through $b_5$.

\section{Fermionic limit}\label{sec:fermionic}
The fermionic limit corresponds to $\alpha'=1-\alpha\to 0$. Technically, the calculations are similar to those for the bosonic case. The expansions of the virial coefficients in $\alpha'$ are as follows:
\begin{align}
&b_2^{\rm anyon} = \phantom{-}\frac14 -\frac12 \alpha'^{\,2},\\[12pt]
&b_3^{\rm anyon} = \phantom{-}
\frac{1}{36} + \frac{1}{12} \alpha'^{\,2},\\[12pt]
&b_4^{\rm anyon} \simeq -0.0149784 \alpha'^{\,2}
\end{align}

Then, to find constants $k_{j\ell}$ one needs to equate coefficients near corresponding powers of $\alpha'$ in expansions of anyonic virial coefficients and virial coefficients of the system under consideration.

At the fermionic end, obviously, 
\begin{equation}
k_{1\ell} \equiv 0.
\end{equation}
and in the second order
\begin{align}
k_{21} \simeq -31.53, \quad
k_{22} \simeq +26.35, \quad
k_{23} \simeq -12.38.
\end{align}
These values yield correct (up to $\alpha^2$) virial coefficients $b_2$ through $b_4$ in the fermionic limit.

The results of calculations are compared to the virial coefficients of anyons in Figs.~\ref{fig:b3Fermi} and \ref{fig:b4Fermi}.

\begin{figure}[h]
\centerline{\includegraphics[scale=0.7]{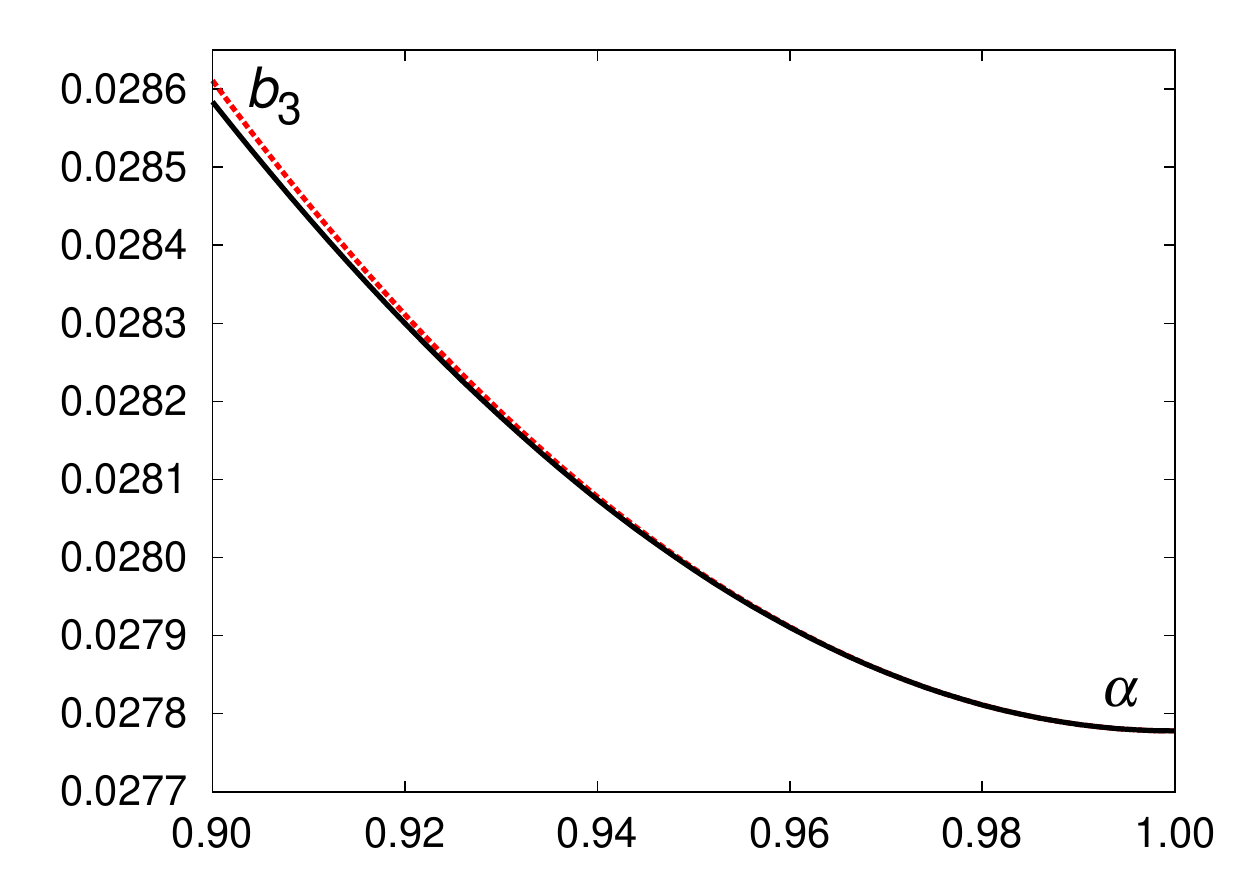}}
\centerline{\includegraphics[scale=0.7]{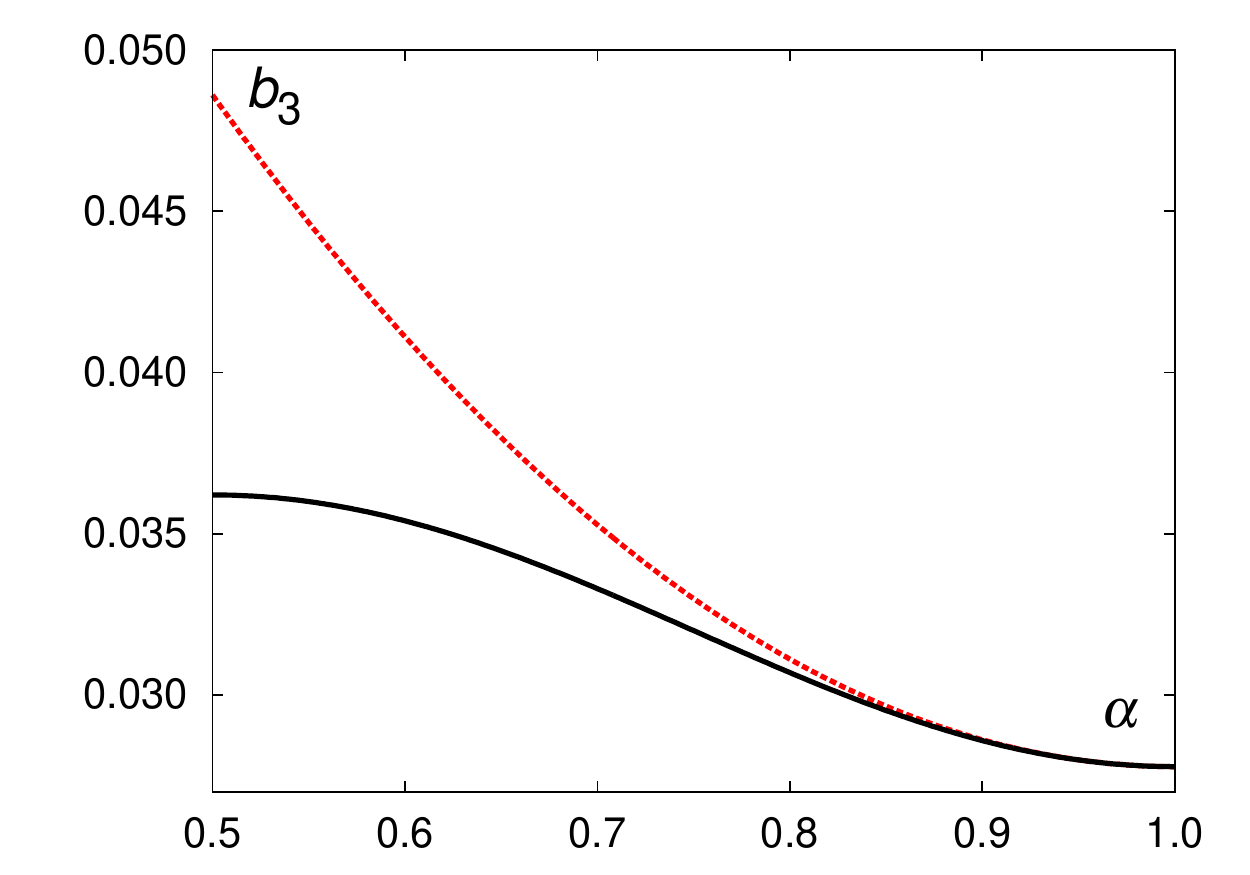}}
\caption{The third virial coefficient:
black solid line --- anyons;
red dotted line --- calculated.
The upper panel corresponds to the close vicinity of the fermionic end while the lower one demonstrates the whole fermionic side $1/2\leq\alpha\leq1$.
}\label{fig:b3Fermi}
\end{figure}

\begin{figure}[h]
\centerline{\includegraphics[scale=0.7]{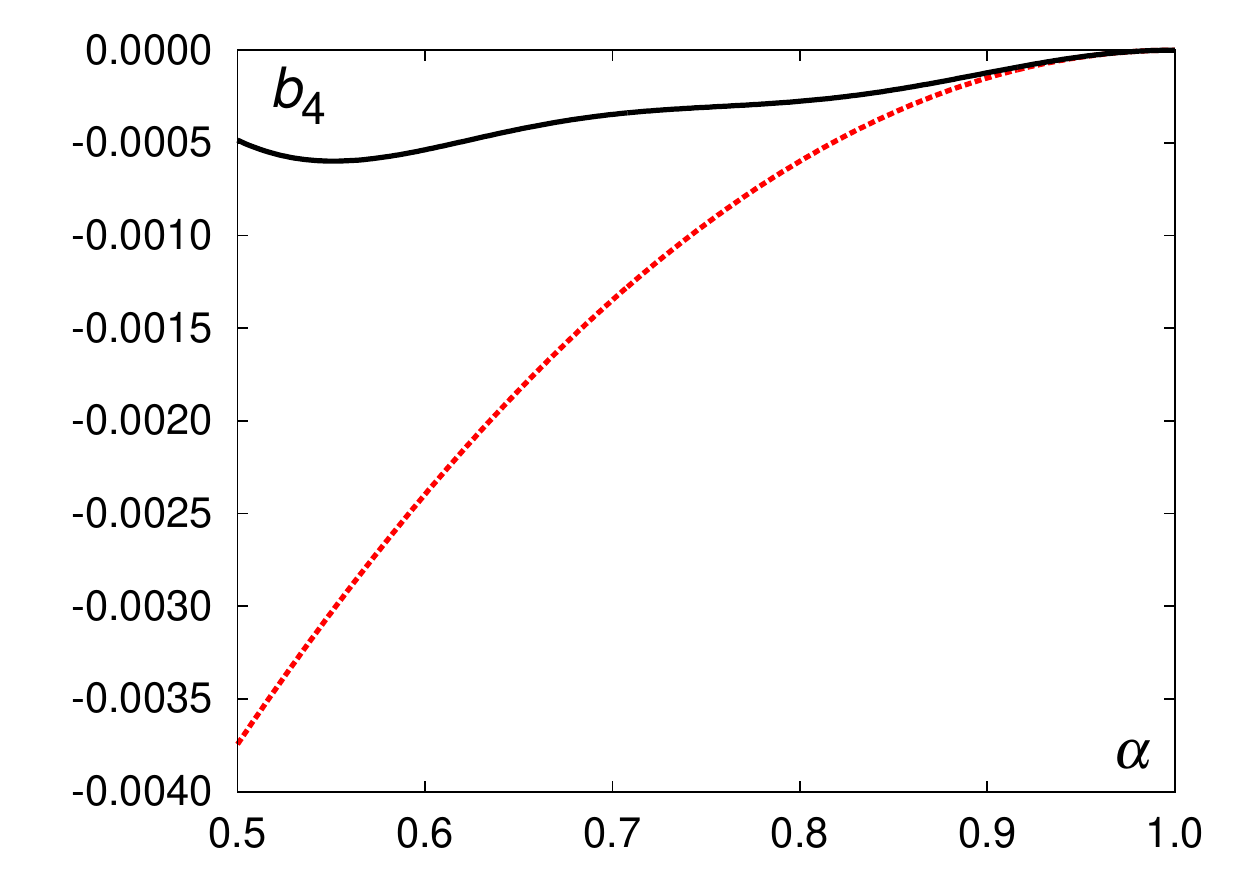}}
\centerline{\includegraphics[scale=0.7]{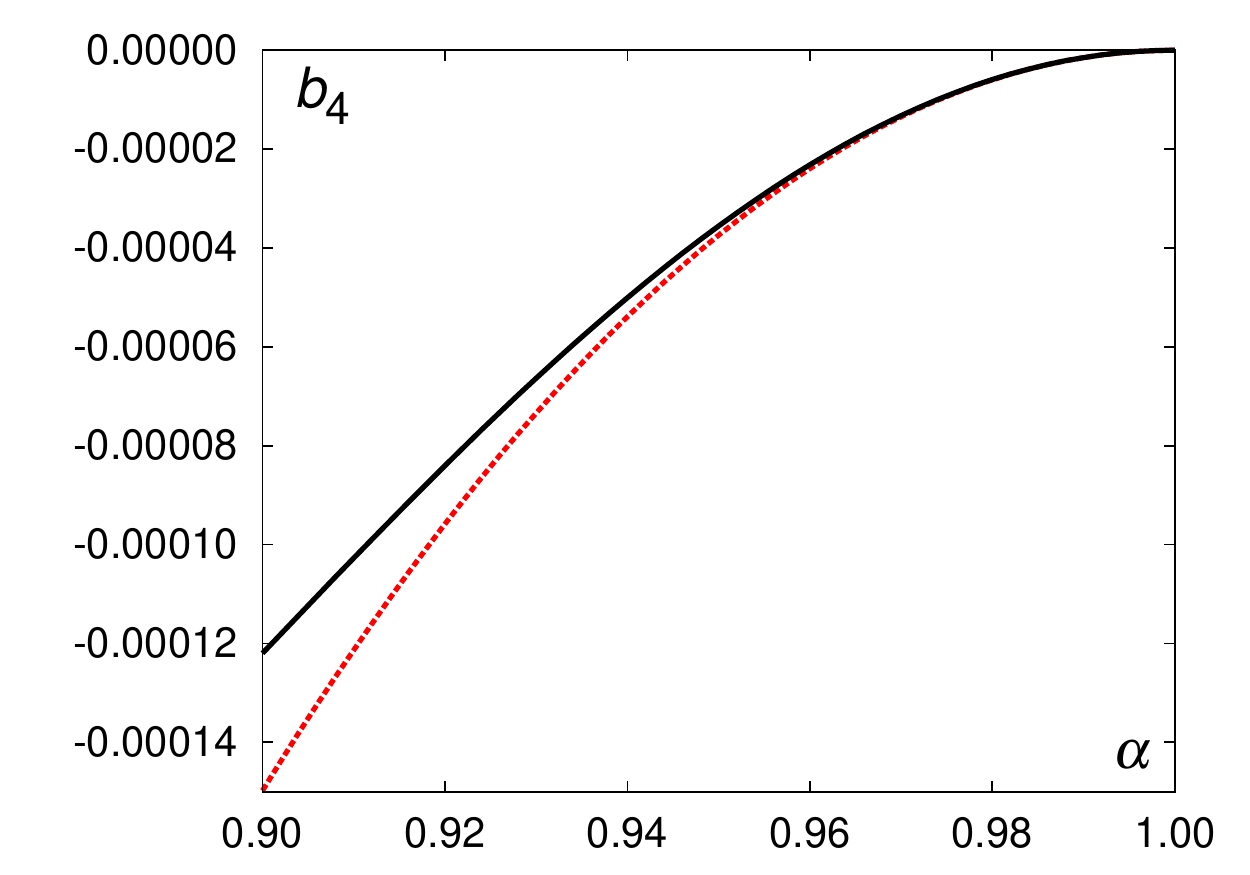}}
\caption{The fourth virial coefficient:
black solid line --- anyons;
red dotted line --- calculated.
The upper panel corresponds to the close vicinity of the fermionic end while the lower one demonstrates the whole fermionic side $1/2\leq\alpha\leq1$.
}\label{fig:b4Fermi}
\end{figure}
\clearpage

\section{Effective excitation spectrum}\label{sec:spectrum}
Instead of deforming the Gibbs factor in the expression for the occupation numbers it is possible to consider the deformed spectrum $\epsilon_{\alpha}$. Indeed, comparing the following relations
\begin{equation}
\exp\left(\frac{\epsilon_\alpha}{T}\right) = 
X\left(\frac{\eps}{T}\right)
\equiv \exp\left(\frac{\eps}{T}\right) 
\left[
1+ \alpha f_1 \left(\frac{\eps}{T}\right) 
+ \alpha^2 f_2 \left(\frac{\eps}{T}\right) 
\right]
\end{equation}
we easily obtain the spectrum in the $\alpha^2$ approximation:
\begin{align}
\epsilon_{\alpha} = \eps\left\{
1+\alpha f_1 \left(\frac{\eps}{T}\right) 
+\alpha^2 \left[ f_2\left(\frac{\eps}{T}\right)- 
\frac12 f_1^2 \left(\frac{\eps}{T}\right)
\right]
\right\}.
\end{align}
Energy $\eps$ corresponds, e.\,g., to $\eps_j=\hbar\omega j$ for harmonically trapped anyons or to $\eps_k=\hbar^2k^2/(2m)$ for free anyons. The correction to the spectrum is shown in Fig.~\ref{fig:spectrum}.

\begin{figure}[h]
\centerline{\includegraphics[scale=1.0]{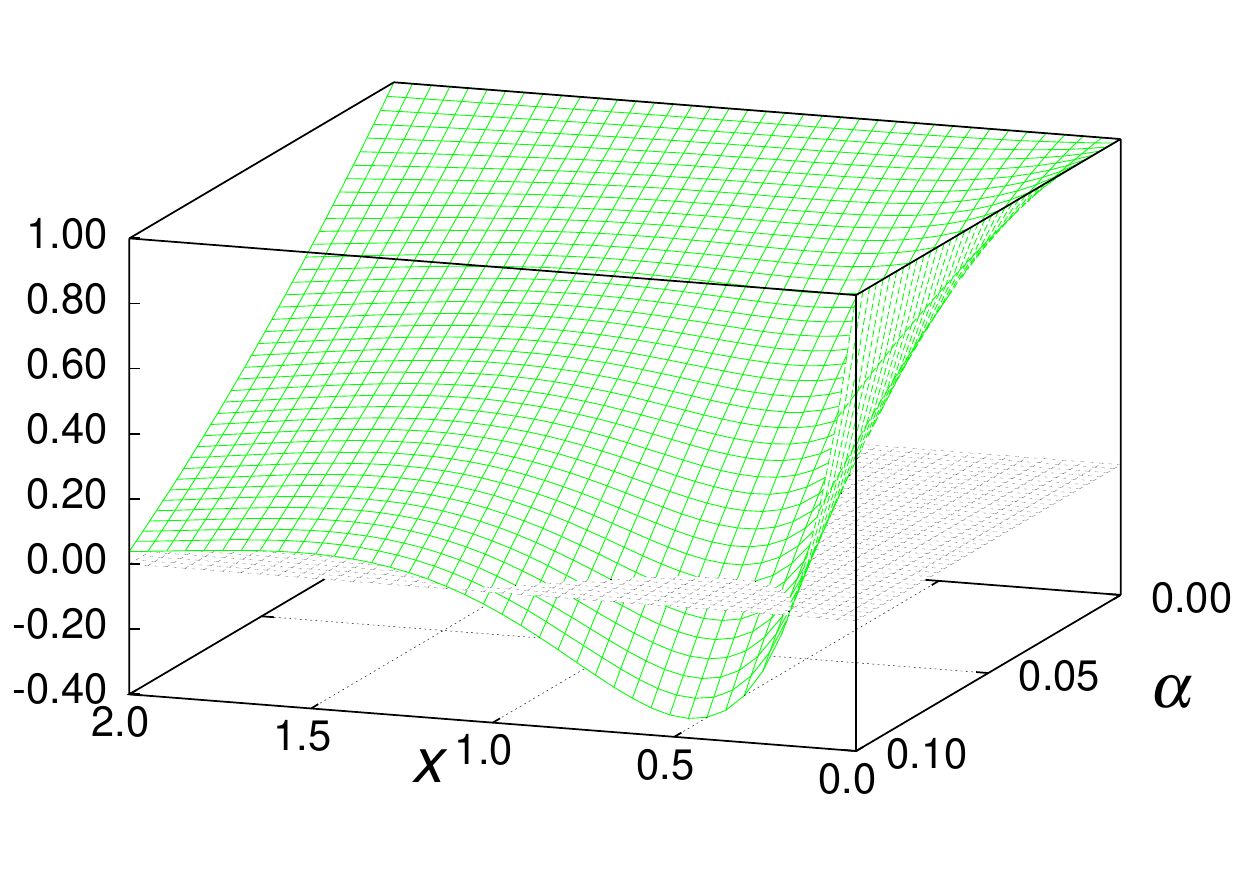}}
\caption{The spectrum correction $\ds\frac{\epsilon_\alpha}{\eps}$ for the anyonic parameter $0\leq\alpha\leq0.1$ and argument $\ds0\leq\frac{\eps}{T}\leq2$.
}\label{fig:spectrum}
\end{figure}

Such an effective spectrum can be used to model anyons using expressions for occupation numbers within the ordinary Bose and Fermi statistics in the appropriate limits.

\section{Discussion}\label{sec:discussion}
We have made an attempt to define a functional form for the occupation numbers of free anyons. Previous models corresponded to exact results for the second and third virial coefficients with a difference only in the fourth virial coefficient leading to a small correction in the equation of state. The approach of the present work reproduces both the fourth and higher virial coefficients in the limits of the Bose and Fermi statistics by means of a deformation of the Gibbs factor in the standard Bose  and Fermi distributions, respectively.

We further plan to juxtapose the obtained statistical-mechanical  description and quantum-mechanical models based on modifications of commutation relations or special algebra for creation--annihilation operators. Extensions of the discussed approach for interaction anyons is also expected.

\section*{Acknowledgment}
This work was partly supported by Project FF-30F (No. 0116U001539) from the Ministry of Education and Science of Ukraine.

\end{document}